\documentclass[prb,twocolumn,showpacs,showkeys]{revtex4}

\usepackage{graphicx}
\usepackage{subfigure}
\usepackage{amssymb}
\usepackage[dvips]{color}

\begin{document}
\title{Ab initio studies of magnetic properties of cobalt and tetracobalt nitride Co$_{4}$N}
\author{Abdesalem Houari}
\affiliation{Theoretical Physics Laboratory, Department of Physics, University of Bejaia, Bejaia, Algeria}
\author{Samir F. Matar}
\affiliation{ICMCB, CNRS,  Universit\'e Bordeaux 1. 87 avenue Dr A Schweitzer,  33600 Pessac Cedex, France}
\author{Mohamed A. Belkhir}
\affiliation{Theoretical Physics Laboratory, Department of Physics, University of Bejaia, Bejaia, Algeria}
\date{\today} 
\begin{abstract}
Magnetic properties and bonding analyses of perovskite structure Co$_{4}$N nitride have been investigated within density functional theory using both pseudo potential and all electron methods. In the same time, the structural and magnetic stability of pure cobalt in hexagonal close packed (HCP), face centered 
cubic (FCC) and body centered cubic (BCC) structures are reviewed. At equilibrium, non-spin polarized (NSP) and spin polarized (SP) calculations of the energy versus volume show that the ground 
state is ferromagnetic in both materials. HCP-Co is found to be more stable than the cubic ones. Magnetic moments of Co atoms in Co$_4$N nitride respectively belonging to two different crystallographic sites are studied over a wide range of the cubic lattice constant, and a comparison with the FCC-cobalt one is given. The volume expansion in the nitride indicates that the corner Co$^{I}$ atoms show localized magnetism while face center Co$^{II}$ atoms exhibit an itinerant behavior. Like in FCC-Fe/Fe$_{4}$N, a ``low volume-low moment'' and ``large volume-high moment'' behavior is observed for FCC-Co/Co$_4$N.  The density of states of the Co$_{4}$N ferromagnetic ground state is interpreted within the rigid band model. The different bonding characters of Co$^I$-N versus Co$^{II}$-N are shown with help of electron localization fucntion ELF plots and spin resolved chemical bonding criteria.
\end{abstract}
\pacs{71.10.-w,71.15.Nc,71.15.Mb}
%
\maketitle 

\section{Introduction}
With respect to their thermal, mechanical and magnetic properties, the transition metal nitrides are known to be very interesting in many technological areas. For example, in the field of high density magnetic recording, iron nitrides are applied [1-5]. Unlike iron nitrides, the cobalt nitrides seem to be less explored and there are only few studies which treat their magnetic properties. The cobalt-nitrogen system was reported to contain many phases ranging from Co$_{4}$N to CoN [6-11]. In the cobalt-rich side, a cubic Co$_{4}$N was synthesised with a lattice constant of 6.7788 bohr (1 bohr=0.529 {\AA}), but no magnetic data were given [7]. On the other side bulk Co$_{2}$N was prepared with an orthorhombic Fe$_{2}$C-type structure [10], whereas no structure was indexed in Co$_{2}$N thin films [11]. However, the case of CoN mononitride seems somewhat conflicting. Like what happens in the iron mononitride [12], it was early reported in cubic rocksalt structure [9], but recently it was found in a paramagnetic zinc-blende type one [13]. On the theoretical side a lack of data about the cobalt nitrides is noted. Except some density functional calculations on CoN [14, 15], no calculations are carried out for Co$_{2}$N and Co$_{4}$N. Nevertheless, for the latter some studies on antiperovskite cobalt-substituted Fe$_{4}$N, i.e. CoFe$_{3}$N, have been undertaken [16, 17].

From the structural point of view, Co$_{4}$N can be assimilated as an insertion of N atom in FCC-cobalt. In this context, a reinvestigation of cobalt metal, at the same time as Co$_{4}$N could be useful for a better understanding of the nitride magnetic properties; especially the Co-N interaction. 

Many experimental and theoretical reports have been carried out in order to study the three known structural types of cobalt, namely body centered cubic (BCC), face centered cubic (FCC) and hexagonal close packed (HCP), see [18-26] and references therein. Under ambient conditions of temperature and pressure, Co is well known to crystallize in hexagonal close packed (HCP) structure. Experimentally a lattice constant in the vicinity of \textit{a}=4.739 bohr, a \textit{c/a} parameters ratio between 1.615\textit{-}1.623 and an atomic magnetic moment of $\sim$1.6\ensuremath{\mu}$_{B}$ were found [18-21]. These results were confirmed later by density functional based calculations [22-23]. Further the stabilization of cobalt in BCC-structure [20], a structural transformation between the two close packed structures: FCC and HCP are also known to occur in cobalt [21]. Recently, the stability of the two latter structures was investigated theoretically but BCC-Co was not included [25]. To our knowledge, the only work which includes the three types of structures is the one of Min \textit{et al} [23] but within the framework of local spin density approximation (LSDA) to the density functional theory [28]. In this context, it is important to remind the well known drawback of this energy functional  in predicting the ferromagnetic BCC-ground state of iron [29] which was of a major impetus in the development of generalized gradient approximation (GGA) [30]. 

The lack of theoretical as well as experimental data concerning  Co$_{4}$N (and in general on the Co-N system), has motivated us to carry out a first principles study. The aim of this present work is to assess the chemical role of nitrogen within Co$_{4}$N and to check whether the  variation for the magnetic moments at different lattice sites follows the trends of cobalt metal in the same FCC-like structure through magnetovolume effects. 

The present paper is organized as follows. In section II the computational framework of calculations and the methods are briefly presented. It is followed by a section, divided in two parts. The first part describes and discusses the phase stability and magnetic moments in cobalt; the second one is devoted to the Co$_{4}$N electronic structure and magnetic properties, where these are treated in details by means of partial density of states and chemical bonding assessments. Finally, section IV presents the  conclusions.

\section{Computational framework and methods}
Two computational methods were used in the framework of density functional theory. 

For a description of the electronic and magnetic structures, the all electrons augmented spherical wave (ASW) method [31] was used. The exchange-correlation effects were accounted for within the generalized gradient approximation (GGA) using the parameterization of Perdew, Burke and Ernzerhof [32]. This scheme was preferred over the LDA (local density approximation) in view of the recent results on iron mono nitrides [12]. In the ASW method, the wave function is expanded in atom-centered augmented spherical waves, which are Hankel functions and numerical solutions of Schr\"odinger's equation, respectively, outside and inside the so-called augmentation spheres. In order to optimize the basis set, additional augmented spherical waves were placed at interstitial sites corresponding to middle of cube edges of the open (in the sense of non close packed) perovskite structure of Co$_{4}$N. The cobalt 4s, 3d and nitrogen 2s, 2p states were treated as valence states. The basis set was complemented by wave functions of s-, p-, and possibly, d-symmetry centered at the interstitial sites which ensure for the covalence effects within the lattice. A sufficiently large number of $k$ points was used to sample the irreducible wedge of the Brillouin zone, and the energy and charge differences between successive iterations were converged below  $\Delta$E=10$^{-8}$ Ryd. and  $\Delta$Q=10$^{-8}$, respectively, in order to obtain accurate values of the magnetic moments and total energy differences between various ordered magnetic states. 

To extract more information about the nature of the interactions between the atomic constituents from electronic structure calculations, the crystal orbital overlap population (COOP) [33] or the crystal 
orbital Hamiltonian population (COHP) [34] may be employed. Both approaches provide a qualitative description of the bonding, nonbonding and antibonding interactions between two atoms. A 
slight refinement of the COHP was recently proposed in form of the 'energy of covalent bond' (ECOV), which combines COHP and COOP to calculate quantities independent of the choice of the zero of potential [35]. Both COOP and ECOV give similar general trends, but COOP, when defined within plane-wave basis sets, exaggerates the magnitude of antibonding states. In the present work the ECOV description was used for the chemical bonding analysis. In the plots, negative, positive and zero ECOV magnitudes are relevant to bonding, antibonding and nonbonding interactions respectively.

In as far as no structural determination is available for Co$_{4}$N trends in cell volumes and ground state crystal structure were desirable in the first place. The equilibrium structures were obtained from a full geometry relaxation of nitride systems starting from Fe$_{4}$N structural setup and using a pseudo potential approach within the VASP code [36]. Ultra soft Vanderbilt pseudo potentials (US-PP) [37] built within GGA scheme [32] were used. The calculations were converged at an energy cut-off of 435.12 eV for the plane-wave basis set with respect to the \textit{k}-point integration with a starting mesh of 6*6*6 up to 12*12*12 for best convergence and relaxation to zero strains. The tetrahedron method with Bl\"{o}chl corrections [38] was applied leading from 141 to 1569 inequivalent tetrahedra. The Brillouin-zone integrals were approximated using a special $k$-point sampling. 

\section{Results and discussions.}
\subsection{Cobalt metal.}
The volume dependence of the total energy for spin polarized (SP) ferromagnetic  and non-spin polarized (NSP -meaning a non magnetic spin degenerate system) configuration for the three cobalt phases: BCC, FCC and HCP are displayed in figure 1-a, 1-b and 1-c, respectively. The ferromagnetic state is largely preferred over the non-magnetic one in all 3 different structures. The lowest energy in the spin polarized configuration is obtained for HCP-Co which is the ambient form of cobalt. This structural stability in the spin-polarized case is better illustrated in figure 1-d, where the E(V) plots for the three structures are shown together in the SP ferromagnetic state. Between HCP and FCC cobalt, a small energy difference \ensuremath{\Delta}E(SP-NSP)$\sim$0.2 eV/atom is obtained; this  agrees well with the low magnitudes reported experimentally and points to the possible coexistence of the two phases at some high pressure range [21]. Furthermore, considering the non magnetic state where FCC cobalt is more stable than HCP-Co, one should conclude that the spin polarization is in favour of the latter.  Our calculations at low volume show that the HCP-phase remains (ferro)magnetic whereas FCC one begins to lose the magnetic character (becomes non magnetic) while being more stable than the former. This finding is in agreement with the recent experimental observations of a non magnetic FCC-phase  preferred to the HCP one in cobalt under high pressure [21].  From figure 1-d, the theoretical equilibrium volumes are nearly equal for HCP and FCC phases ($\sim$78.27 bohr$^3$.atom$^{-1}$), whereas the BCC one is slightly larger ($\sim$79.51 bohr$^3$.atom$^{-1}$). In terms of total energy, and as confirmed by recent theoretical study [26], a BCC-cobalt phase as bulk material should not be stable at ambient pressure. Nevertheless, the experimentally stabilized BCC cobalt thin films [20] were explained to originate from such imperfections or substrate lattice constant mismatch at an early stage of growth. Furthermore a critical thickness above which the BCC phase begins to transform to HCP one was identified [27]. In HCP-cobalt, our c/a ratio (1.603) and magnetic moment (1.634 $\mu_B$) are in good agreement with experiments [18-21] and in better agreement than recent full potential calculations [25]. For FCC cobalt no experimental report is available for the cobalt magnetic moment; on the other hand our computed value of  $\sim$1.67 $\mu_B$  is  in good agreement of theoretical results both in LSDA [23] and GGA [25] exchange-correlation functionals. As for BCC-Co, it needs to be stressed that our 1.757 $\mu_B$ moment magnitude which agrees with computed LSDA [23] one, is larger than the only reported experimental value of 1.53 $\mu_B$ [20]. 

\subsection{Tetracobalt nitride Co$_4$N.}
\subsubsection{Geometry optimization and electron localization from pseudo potential calculations.}
We started the geometry optimization by using Fe$_{4}$N structural setup with a lattice parameter from experiment: a=7.165 bohr. We recall that this structure  can be viewed from a cubic perovskite lattice: ABX$_{3}$ where A is a large cation at corner, B is a smaller  cation at cube center and X is an anion at face centers. From this  $Co^{I}Co^{II}_{3}N$ is an anti-perovskite because Co$^{II}$ replace X anion sites at face centers. Then the lattice geometry was fully relaxed for the cell volume, shape and atomic positions. The calculations were carried out for both non magnetic and spin polarized configurations. The resulting structure was found to be cubic with lattice constants: a$_{NSP}$(Co$_{4}$N)=6.975 bohr, a$_{SP}$(Co$_{4}$N)=7.062 bohr, with an energy difference of \ensuremath{\Delta}E(SP-NSP)= -0.982 eV favoring the spin polarized configuration. The experimental value of a$_{exp.}$(Co$_{4}$N)= 6.779 bohr [7] is significantly smaller and this calls for an interpretation.

One explanation can be found in the existence of N vacancies within the lattice. A geometry optimization with a nitrogen deficient system was carried out calling for a new cell set up. Starting from a supercell with 8 formula units leading to  32 Co and 8 N, the above calculated lattice constant of Co$_4$N was checked firstly. Then vacancies were created at nitrogen sites. With 2 N vacancies, a small change of lattice constant is obtained with a= 6.925 bohr; with 4 nitrogen vacancies the calculated lattice constant becomes smaller, with a = 6.762 bohr. This magnitude comes close to the experimental value. These results support the potentiality of a nitrogen deficient lattice in the sense of the existence of many nitrogen lacking  sub-stoichiometric domains (Co$_{4}$N$_{1-x}$, x$\sim$0.5), beside stoichiometric Co$_{4}$N. While such a result is not reported experimentally [7], we suggest that nitriding conditions calling for high nitrogen pressures can be useful to obtain the stoichiometric compound [39].
Although the supercell calculation results allowed to clarify the hypothesis of nitrogen vacancies with the tetracobalt nitride lattice, it should be kept in mind that they assumed an ideal bulk lattice with an ordered distribution of nitrogen atoms and vacancies, while in experimental reality those should be randomly distributed with possible surface effects. 

A quantity that can be extracted from calculations is the electron localization function ELF which allows determining the amount of localization of electrons with respect to the free electron gas distribution as described by Becke and Edgecomb [40]. Figure \ref{fig2} shows a $<101>$ diagonal plane crossing all atomic sites of Co$_{4}$N.  This illustration stresses the isolated character of the corner sites Co$^{I}$ at a\ensuremath{\sqrt{3}}/2 from N with respect to face center cobalt Co$^{II}$ which are at a/2 from N. Further there is a non negligible localization along Co$^{II}$-N-Co$^{II}$  direction. Nitrogen exhibits the largest localization (large white area around it) in agreement with the description of the system as a ``nitride'' from the chemical view point. On the contrary, the metal sites  show no localization (dark areas) due to their electropositive character so that electrons are smeared out in between the lattice sites (light grey areas) like in a metal, which is the characteristics of the system under study. 

\subsection{Analysis of the electronic and magnetic structure}
In this section, we investigate in detail the magnetic and electronic properties of pure Co$_{4}$N. The chemical effects of nitrogen are studied by means of partial density of states (PDOS) and chemical bonding 
properties (covalence energy ECOV). Within FCC cobalt the result of nitrogen insertion is to redistribute the FCC lattice into two different sublattices: Co$^{I}$ (cube corners) and Co$^{II}$ (face centers). The same atomic arrangement is found within Fe$_{4}$N but intriguingly the lattice parameter of the pure Co$_{4}$N found experimentally (\textit{a}= 6.7788 bohr) [7] is much smaller than the Fe$_{4}$N one (\textit{a}= 7.1777 bohr) [22]. Moreover, Matar \textit{et al} [16] have carried out an experimental study, reinforced by theoretical calculations, on cobalt-substituted Fe$_{4}$N and have shown that Co can substitute Fe in both corners (CoFe$_{3}$N) and face centers sites (FeCoFe$_{2}$N) with no change in Fe$_{4}$N lattice constant value. Furthermore, considering the iron and cobalt atomic radii (Co is slightly smaller than Fe: r$_{Fe}$=2.212 bohr and r$_{Co}$=2.193 bohr [16]), one can conclude that Fe$_{4}$N and Co$_{4}$N must have nearly the same magnitude of lattice constant. As commented in the previous section and demonstrated by geometry optimization for N deficient system, this may be due to the presence of N vacancies in the elaborated samples. The  theoretical lattice parameter of the system in the magnetically ordered state, which is very close to the  Fe$_{4}$N one [22], agrees well with the experimental study of Co-substituted Fe$_{4}$N [16]. However it is larger than the value reported for Co$_{4}$N [7]. 

The values of total energy at equilibrium show that the ferromagnetic state is preferred over the non magnetic 
one with \ensuremath{\Delta}E(SP-NSP)= -1.4 eV per fu. This value is somehow close to the energy stabilization of SP ground state from US-PP calculations because the latter method accounts for valence electrons only. A magnetic volume expansion is visible whereby the theoretical lattice constant increases from 7.010 bohr in the non magnetic state at equilibrium to 7.098 bohr in the ferromagnetic state so that one expects the system to be more compressible in the spin polarized configuration. In the latter, we have obtained magnetic moment values of 1.967\ensuremath{\mu}$_{B}$, 1.486\ensuremath{\mu}$_{B}$ and 0.069\ensuremath{\mu}$_{B}$ for Co$^{I}$, Co$^{II}$ and N respectively. With respect to the average value of 1.6 $\mu_B$ of Co, the enhancement of the magnetic moment  of corner Co$^{I}$ underlines its isolated character on one hand and the reduction of Co$^{II}$ moment points to the chemical role of neighboring nitrogen on the other hand. Experimentally no reports about Co$_{4}$N magnetic moments are available. Compared to the data on CoFe$_{3}$N, our magnetic moments agree with reference [16] for cobalt Co$^{I}$ (situated in the corners) and for nitrogen; whereas a large difference exists for Co$^{II}$. This fact is expected because the previous authors have substituted only one of the 
three face centers irons, and thus their Co$^{II}$ moments (0.842\ensuremath{\mu}$_{B}$, 
see [16]), are largely affected by the neighbouring irons. Nevertheless, our Co$^{I}$ moment is smaller than 2.26\ensuremath{\mu}$_{B}$ computed in ref. [17] for CoFe$_{3}$N. 

\subsubsection{Comparison of FCC-Co versus Co$_{4}$N}
The magnetovolume effect of nitrogen insertion within Co$_{4}$N is illustrated in figure \ref{fig3}. Moreover, since Co$_{4}$N can be seen as FCC-Co with N atom at the body center, 
we have included the variation of the moment in FCC-Co in order to allow a better check of this effect. One can distinguish two different parts in figure \ref{fig3}. At low volume (below FCC-Co equilibrium), the Co$^{I}$ moment rises rapidly to high value reaching 1.909\ensuremath{\mu}$_{B}$ at the FCC-Co theoretical equilibrium volume (just below the Co$_{4}$N one); while Co$^{II}$ moment increases somewhat slowly and a value of 1.486\ensuremath{\mu}$_{B}$ is reached at equilibrium. Beyond the Co$_{4}$N 
equilibrium volume, the Co$^{II}$ moment still increases while Co$^{I}$ one seems to stabilize at high volume. This variation can be assimilated to the moment variation of FCC-Co. At low volume the latter resembles Co$^{I}$ and increases rapidly to a high value with increasing volume. However, beyond its theoretical equilibrium the FCC-Co moment varies like Co$^{II}$. As a consequence it can be concluded that the face centers Co atoms (Co$^{II}$) show itinerant magnetism while the corners ones (Co$^{I}$) exhibit a localized behavior. The observed behavior can be explained in terms of Co-N bond \textit{p-d} hybridization in the two different cobalt sites Co$^{I}$ and Co$^{II}$. While Co$^{I}$ is far from the cube center, it seems almost not affected by the nitrogen atom and shows a great sensibility to the onset of volume variation by acquiring a large moment value and it begins to stabilize only at high volume. On the contrary, Co$^{II}$ is subjected to the chemical presence of neighboring N atoms at a/2 and 
the strong \textit{p-d} hybridization makes its moment vary slowly with the volume until a sufficient value (beyond the FCC-Co equilibrium) where its variation begins to be more pronounced. From this, it can be concluded that the two kinds of moment variation in FCC-Co are simultaneously present in Co$_{4}$N and this is due to the insertion of N atom. We note finally that this behavior is observed, with almost the same manner as in Fe$_{4}$N versus FCC-Fe [41,42]. 

\subsubsection{Analyses of the density of states and bonding properties}

\begin{figure}[htbp]
\begin{center}
\subfigure[~]{\includegraphics[width=0.45\linewidth]{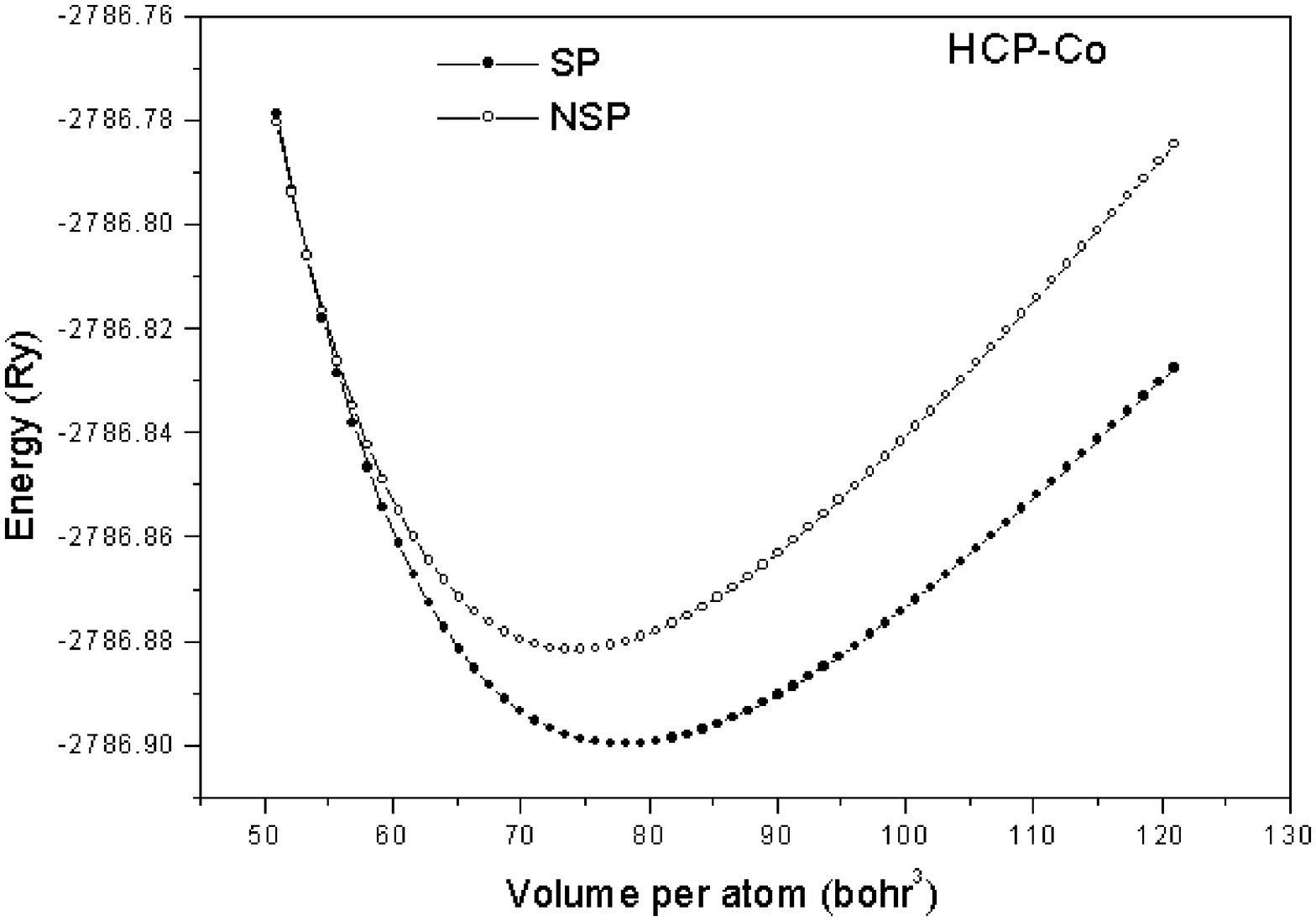}}
\subfigure[~]{\includegraphics[width=0.45\linewidth]{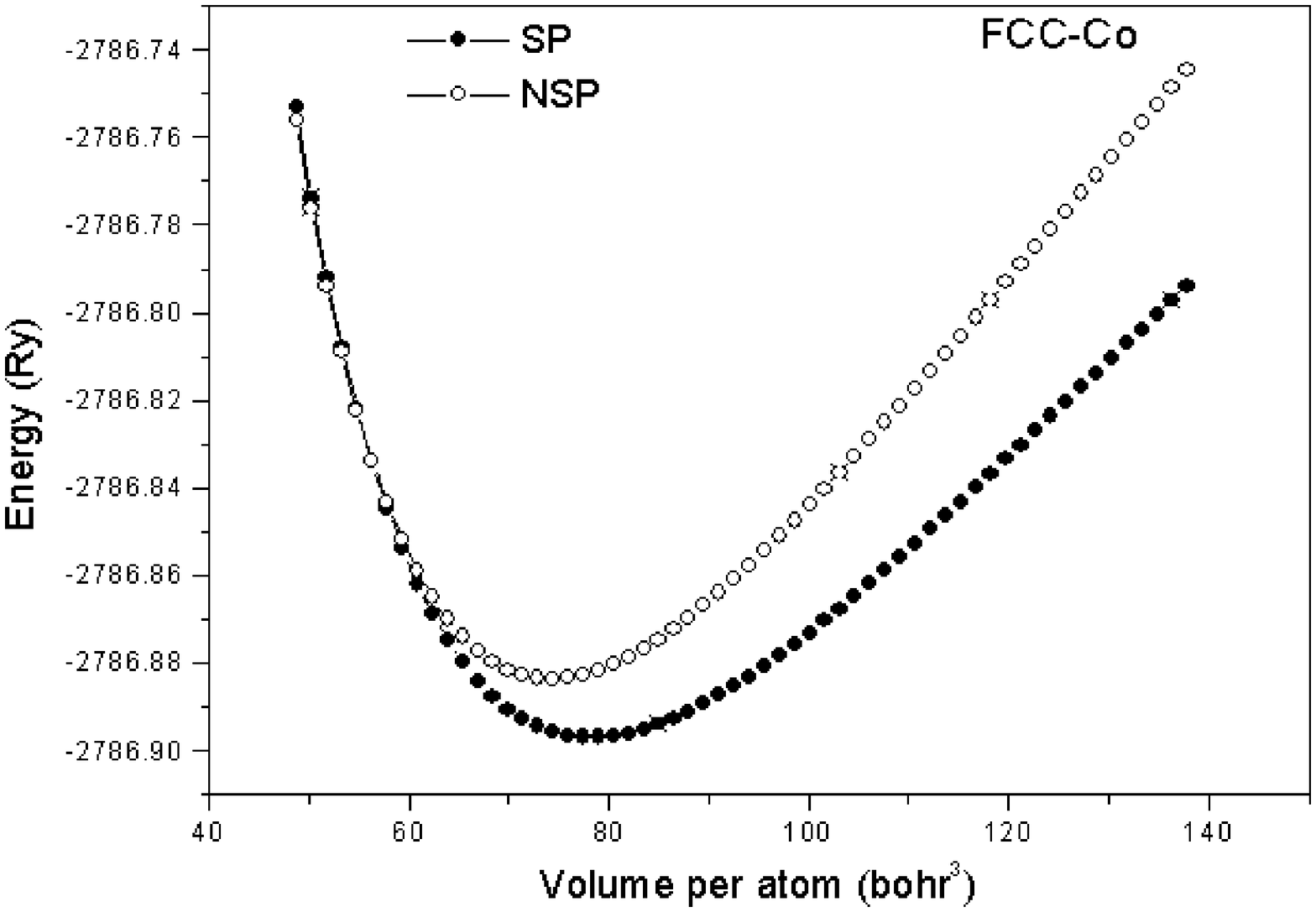}}
\subfigure[~]{\includegraphics[width=0.45\linewidth]{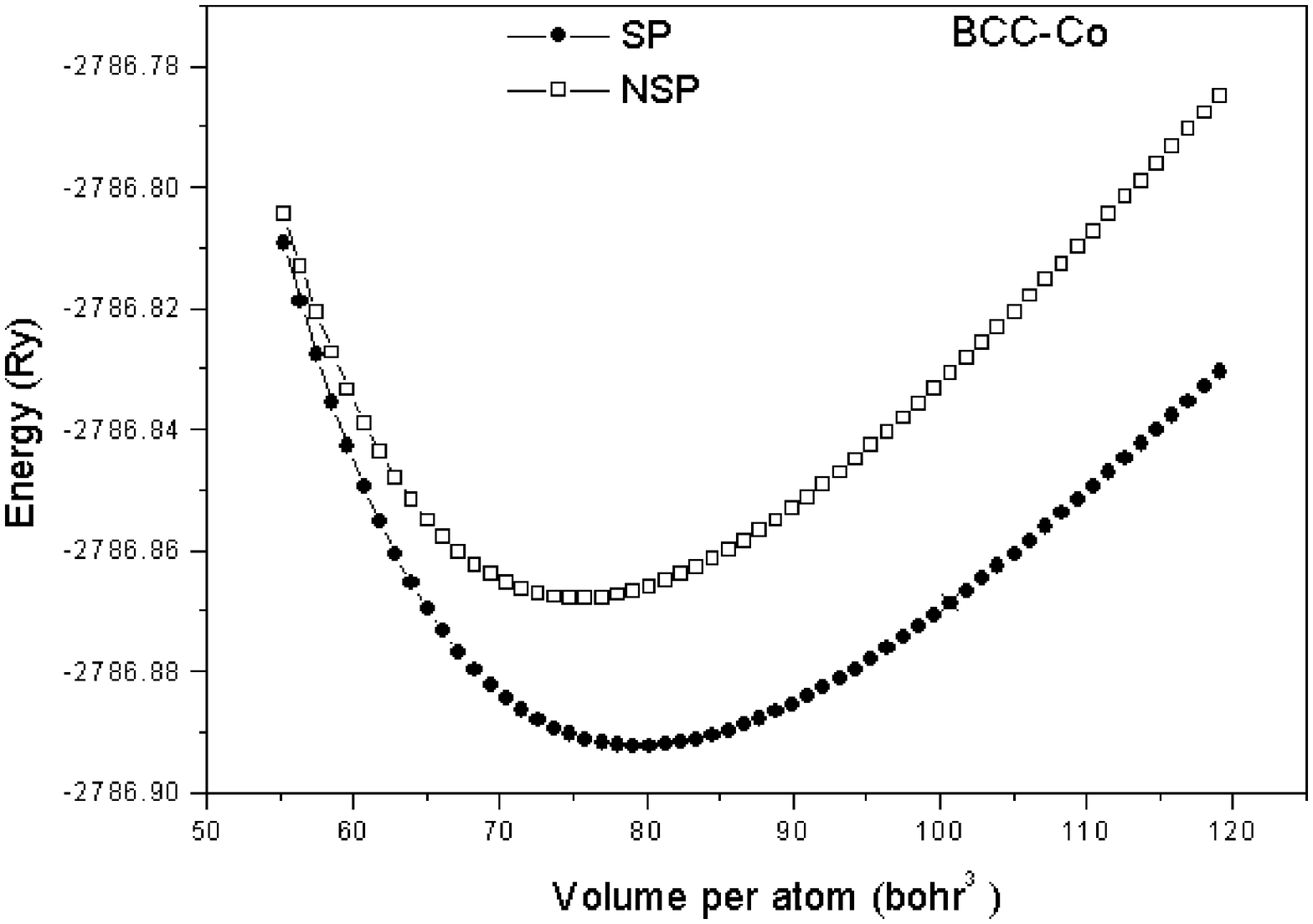}}
\subfigure[~]{\includegraphics[width=0.45\linewidth]{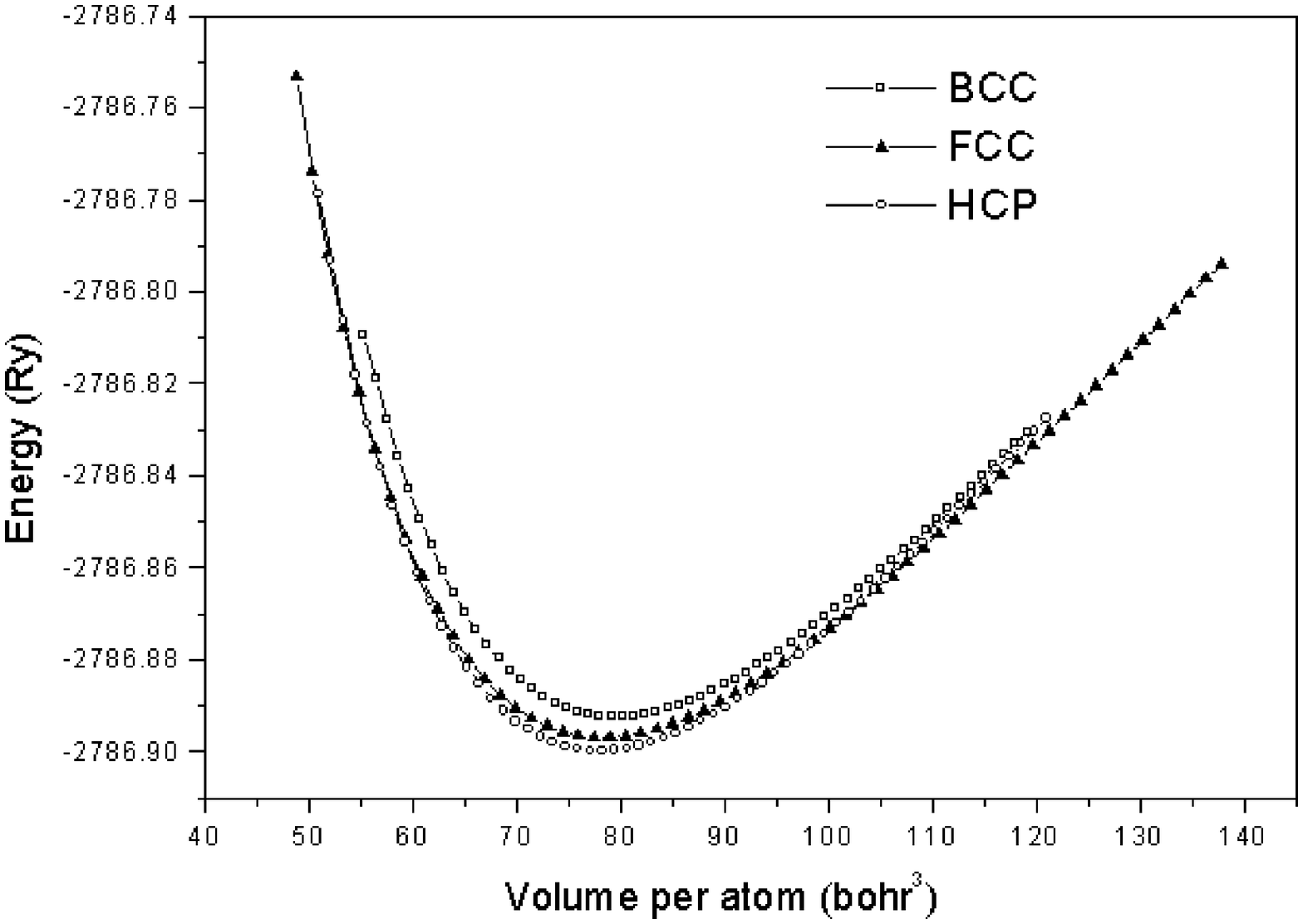}}
\caption{Energy versus volume curves for non spin polarized (NSP ) and spin polarized (SP) cobalt in three allotropic varieties a) HCP, b) FCC, c) BCC and d) comparison of the three structures of Co in ferromagnetic ground state.}
\label{fig1}
\end{center}
\end{figure}

\begin{figure}[htbp]
\begin{center}
\subfigure{\includegraphics[width=0.76\linewidth]{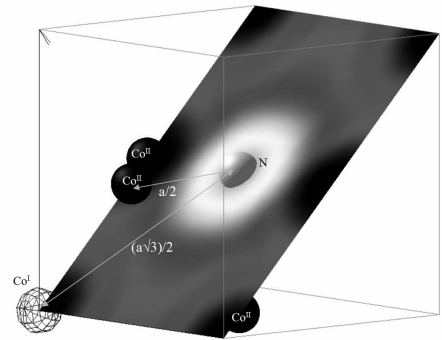}}
\caption{Sketch of the anti perovskite cubic structure of Co$_{4}$N ($Co^ICo^{II}_3N$) with a diagonal 101 plane showing electron localization function (ELF) contour plots.  Strong localization (around N), free elecron gas like distribution (within the lattice) and no localization (around cobalt sites) correspond to white, grey and black areas respectively.}
\label{fig2}
\end{center}
\end{figure}

\begin{figure}[htbp]
\begin{center}
\subfigure{\includegraphics[width=0.8\linewidth]{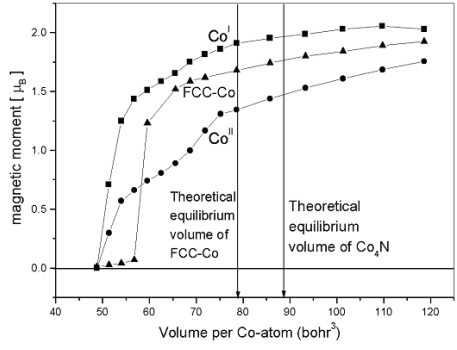}}
\caption{Magnetic moment versus volume in Co$_{4}$N and FCC-Co.}
\label{fig3}
\end{center}
\end{figure}

\begin{figure}[htbp]
\begin{center}
\includegraphics[width=0.8\linewidth]{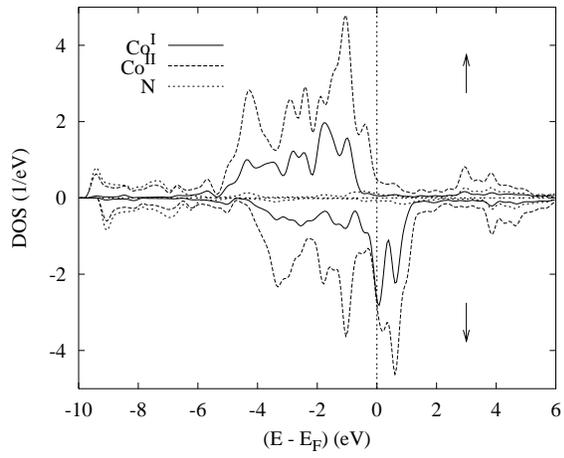}
\caption{Site and spin projected density of states (DOS) of spin polarized (SP) Co$_{4}$N.}
\label{fig4}
\end{center}
\end{figure}

\begin{figure}[htbp]
\begin{center}
\subfigure[~]{\includegraphics[width=0.5\linewidth]{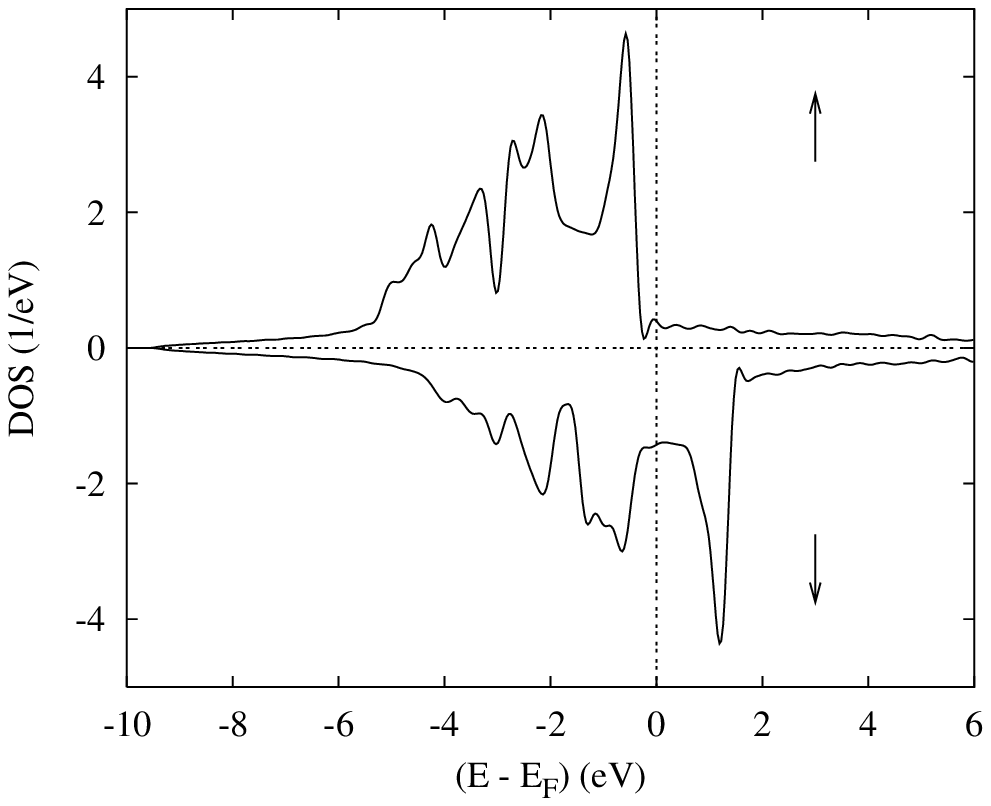}}
\subfigure[~]{\includegraphics[width=0.5\linewidth]{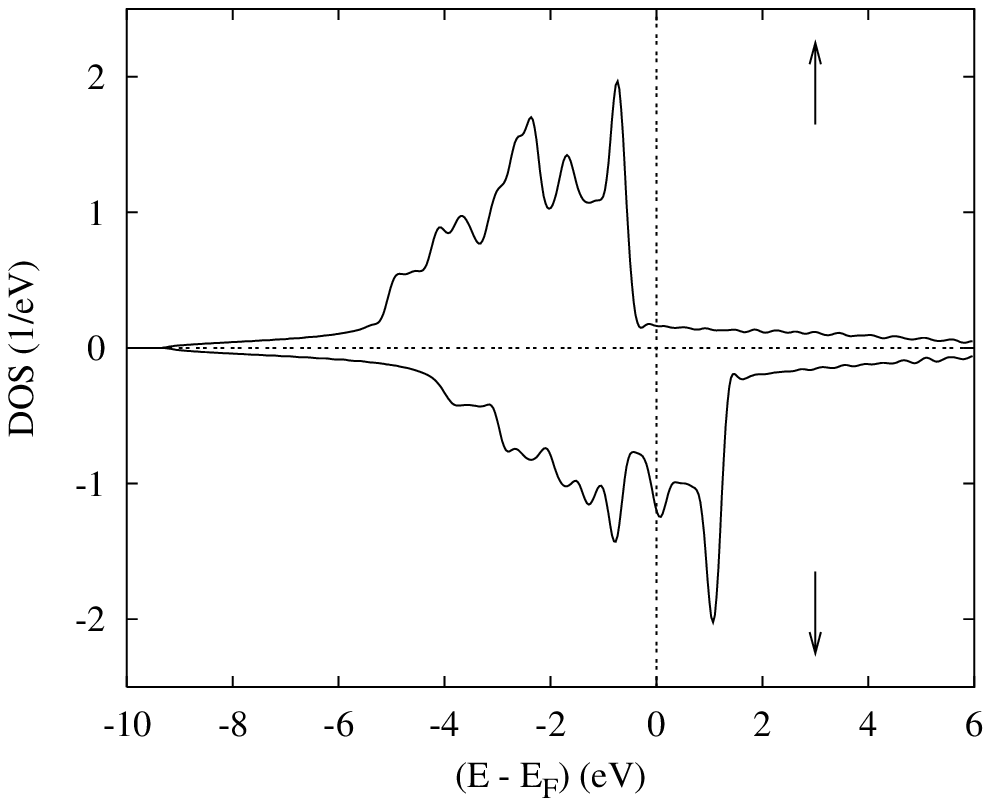}}
\subfigure[~]{\includegraphics[width=0.5\linewidth]{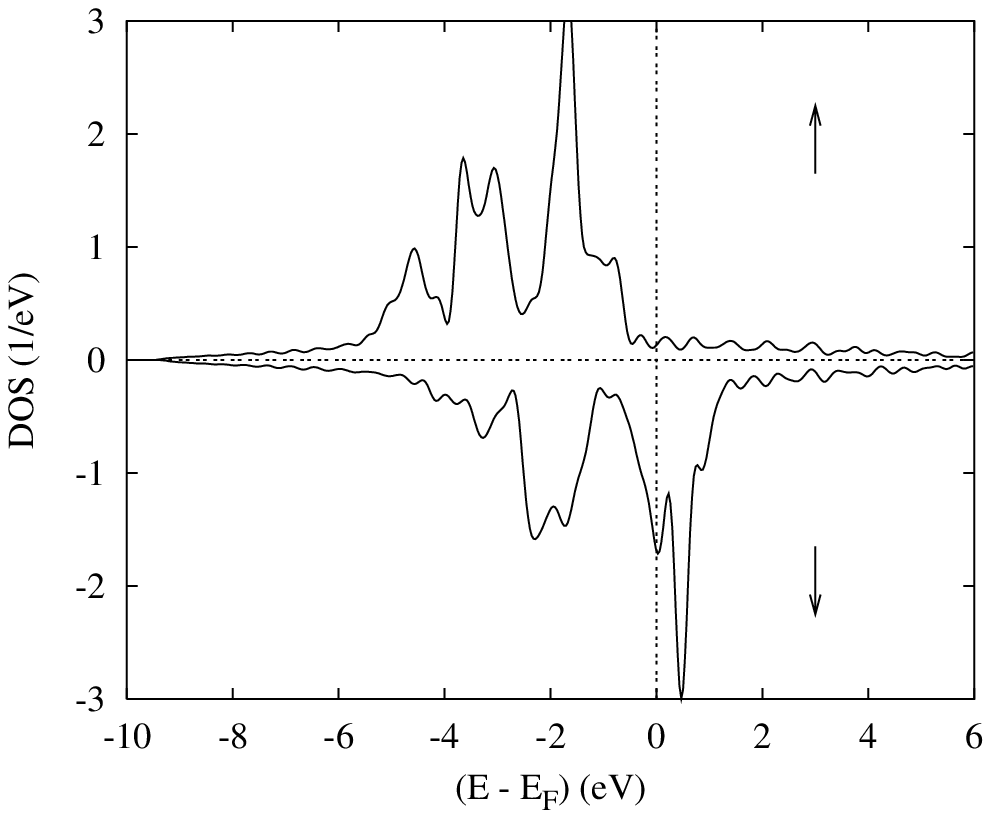}}
\caption{DOS of spin polarized (SP) cobalt in three allotropic varieties (a) HCP, (b) FCC, (c) BCC .}
\label{fig5}
\end{center}
\end{figure}

Using equilibrium values of lattice constants determined above, we examine the spin ($\uparrow$ and $\downarrow$)  and site projected density of states (PDOS) accounting for site multiplicity in the formula unit: 1Co$^{I}$, 3Co$^{II}$ and 1N. This is shown in figure \ref{fig4}. Further, for sake of comparison  we present the DOS of three studied varieties of Co in figure \ref{fig5} at equilibrium lattice volumes (cf. fig. \ref{fig1}). From the very low spin $\uparrow$ DOS at the Fermi level due to the filling of the Co(3d$^\uparrow$) sub band up to 5 electrons, the major feature of the latter is its strong ferromagnetic behavior in all three varieties; we note the resemblence of  BCC-Co  DOS  with BCC-Fe DOS \cite{43}. Thus from the point of view of strength of magnetic behavior cobalt is like FCC-Ni whose pressure for the disappearance of magnetic moment is $\sim$4000 kbar, contrary to BCC-Fe which is a weak ferromagnet \cite{43}. 

In Co$_4$N nitride, the energy shift between spin $\uparrow$ and $\downarrow$ populations, due to the exchange splitting is larger for Co$^{I}$ than for Co$^{II}$. From Hankel spherical wave function energies E$_{Hankel}$ which, in the ASW formalism, are relative to the middle of the band, we compute the following exchange splitting magnitudes: $\Delta E_{Hankel}(Co^{I})$=2.74 eV and $\Delta E_{Hankel}(Co^{II})$=1.42 eV while an intermediate value of $\Delta E_{Hankel}(Co)$= 1.66 eV is obtained for FCC Co. The magnitudes of spin splitting are proportional to the magnetic moments. They  can be interpreted by the more localized PDOS of Co$^{I}$ as with respect to Co$^{II}$ atoms which undergo the chemical pressure from nitrogen with a subsequent lower magnetic moment. Due to the electronegative character of nitrogen its PDOS are  found at lowest  energy within the VB, between -10 and -6 eV. One common feature between tetracobalt nitride and FCC-Co  is the very low DOS at the Fermi level for majority spin which means that both Co sites in the nitride contribute to make the system behave as a strong ferromagnet just like FCC-Co. Particularly one can notice the low intensity peak at \ensuremath{\sim}0.5 eV below the Fermi level in the spin \ensuremath{\uparrow} panel which arises from the Co$^{II}$-N \ensuremath{\pi}* bonding. This is a major difference with Fe$_{4}$N [41,42] where the corresponding peak due to Fe$^{II}$-N \ensuremath{\pi}* lies exactly at the Fermi level and contributes to assign a weak ferromagnetic behavior for the iron based nitride. 

\begin{figure}[htbp]
\begin{center}
\subfigure[~]{\includegraphics[width=0.7\linewidth]{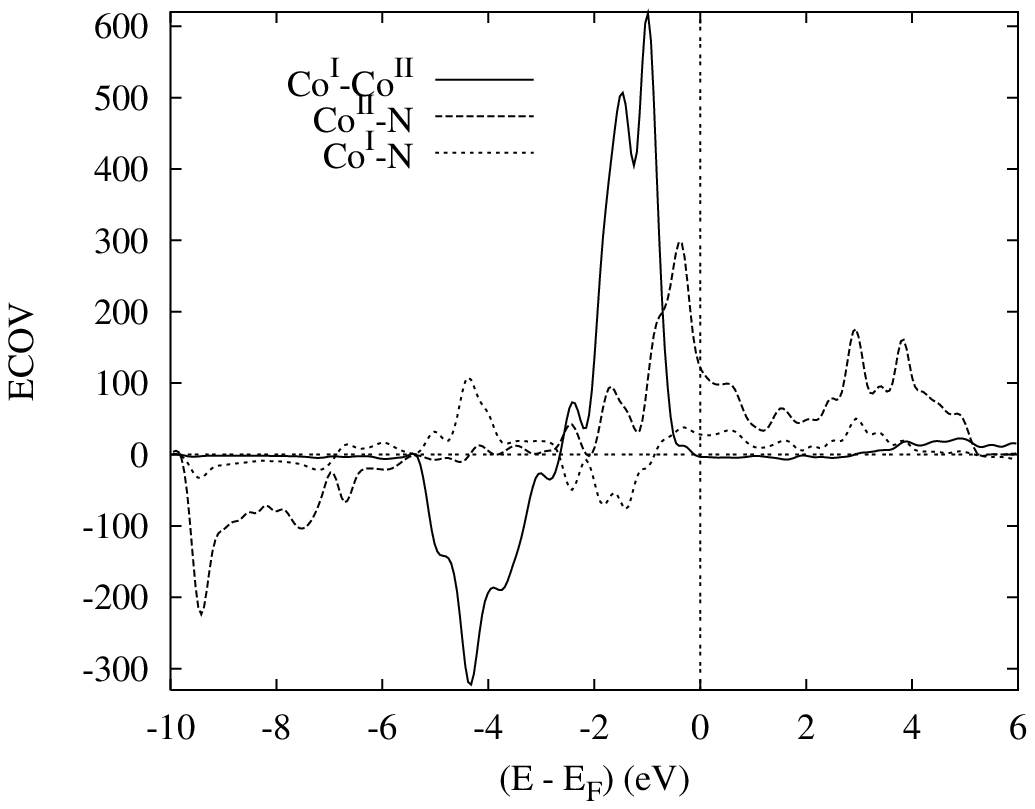}}
\subfigure[~]{\includegraphics[width=0.7\linewidth]{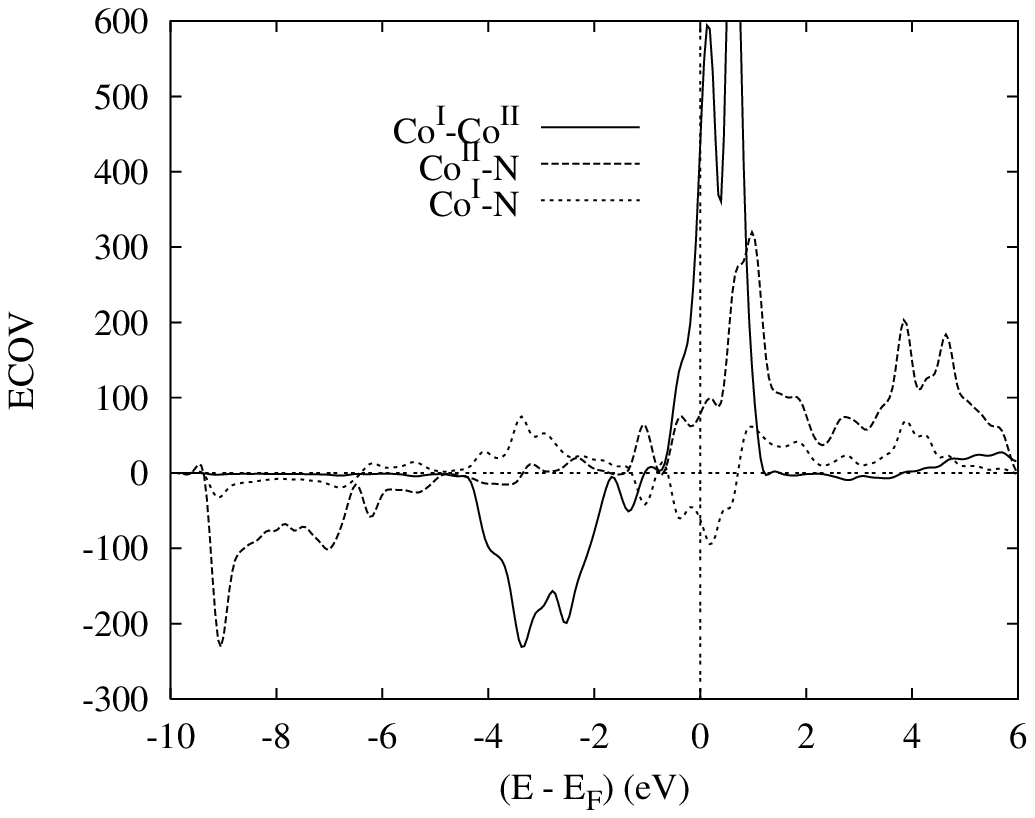}}
\caption{Spin resolved chemical bonding from ECOV criterion within Co$_{4}$N: (a) majority spins ($\uparrow$), (b) minority spins ($\downarrow$).}
\label{fig6}
\end{center}
\end{figure}

For further explanations of the quantum mixing between the valence states, the two atom chemical interactions are examined within the covalent bond energy ECOV plotted for the spin polarized ferromagnetic  state in figure \ref{fig6}. Negative and positive ECOV magnitudes designate bonding and antibonding interactions respectively. Looking firstly at the bonding within the Co metal sublattices, due to the large filling of Co d states the Co$^{I}$-Co$^{II}$ interaction is found half bonding/half antibonding. The exchange splitting causes this interaction to stabilize for majority  ($\uparrow$) spin at lower energy so that it is found within the VB  (panel \ref{fig5}-a) while its destabilization to higher energy for minority  $\downarrow$ spin (panel \ref{fig5}-b) places the antibonding states on the Fermi level. This means that $\downarrow$ Co$^{I}$-Co$^{II}$ interaction is more bonding than $\uparrow$ one. The  bonding with nitrogen is clearly dominant for face center cobalt and is found mainly in the lower part of the VB with $\sigma$ like bonding sharp peaks at $\sim$-9 eV and more extended $\pi$-like bonding in the [-9, -6 eV] range. They are followed by antibonding counterparts  $\pi$* around E$_{F}$ as suggested in the DOS discussion and $\sigma$* above E$_{F}$ within the conduction band at $\sim$~4 eV. From the relative energy position between the two panels of fig. \ref{fig6}, Co$^{I}$-N and Co$^{II}$-N are little  affected by the exchange splitting and they are both dominantly bonding within the VB so that nitrogen plays a role in stabilizing Co$_{4}$N nitride. 

\section{Conclusions.} 
In this work, we have studied the electronic and magnetic properties of Co$_{4}$N compound with respect to pure cobalt in three different structures: HCP, FCC and BCC. For the latter, the obtained results 
were found in good agreement with experimental and theoretical ones. Such a reinvestigation, especially for 
FCC-cobalt was very useful to understand the magnetovolume effects in Co$_{4}$N. We showed that the magnetic moment variation in the latter could be assimilated to FCC-Co. The low volume (low moment) 
and large volume (high moment) states of FCC-cobalt are found to be present, at the same time, in the two cobalt sites Co$^{II}$ and Co$^{I}$ of Co$_{4}$N. The Co$^{I}$ moment appears very sensitive to 
the onset of volume variation, like cobalt at low volume, and reaches saturation rapidly, whereas the Co$^{II}$ one begins to increase significantly only beyond equilibrium and without fully saturating. 
This behavior (low volume-low moment and large volume-high moment) is explained, via the density of states and chemical bonding analyses, in terms of Co-N relative distances and the corresponding 
d-p hybridization. The Co$^{I}$ atom at a${\sqrt{3}}$/2 from N atom is not affected and is found rather localized while the Co$^{II}$ one which is at only a/2 from N is submitted to a \textit{chemical 
pressure} due to nitrogen and has itinerant behavior. Although similarities with Fe$_{4}$N can be traced out, the overall DOS features point to a strong ferromagnetic nitride for both Co sites. 

\section{Acknowledgements} 
Computations were partly carried out on main frame computers of the M3PEC-M\'esocentre - University Bordeaux 1 Science and Technology (http://wwww.m3pec.u-bordeaux1.fr), financed by the ``Conseil R\'egional d'Aquitaine'' and the French Ministry of Research and Tecnology.

{}

\newpage

\begin{thebibliography}{}
\bibitem {1} D. Andriamandroso, G.Demazeau, M. Pouchard and P. Hagenmuller, J. Solid State Chem.\textbf{54}, 34  (1984). 
\bibitem{2} S. Matar, P. Mohn, G. Demazeau and B. Siberchicot, J. Phys. France \textbf{49}, 1761 (1988). 
\bibitem {3} S. Ishida  and K. Kitawatase, J. Magn. Magn. Mater. \textbf{130}, 353 (1994).
\bibitem {4} A. Sakuma, J. Magn. Magn. Mater. \textbf{102}, 127 (1991).
\bibitem {5} R. Coehoorn, G.H.O. Daalderop  and H.F.J. Jansen, Phys. Rev. B \textbf{48}, 3830 (1993).
\bibitem {6} M. Morito, O. Kenichi and I. Takashi, Appl. Phys. Lett \textbf{49}, 977 (1986).
\bibitem {7} Kiichi Oda, Tetsuo Yoshio and Kohei Oda, J. Mater. Science \textbf{22}, 2729 (1987).
\bibitem {8} M. Beshkova, G. Beshkov, M. Marinov, D. Bogdanov-Dimitrov, G. Mladenov, T. Tanaka and K.Kawabata , Mat. Manuf. Process \textbf{16}, 531 (2001).
\bibitem {9} O. Schmidt-Dumont and N. Kron, Angew. Chem. \textbf{67}, 231 (1955).
\bibitem {10} M. Hasegawa and T. Yagi, Sol. Stat. Comm \textbf{135}, 294  (2005). 
\bibitem {11} W. De la cruz, O. Contreras, G. Soto and E. Perez-Tijerina, Rev. Mex. Phys \textbf{52}, 409  (2006).
\bibitem {12} A. Houari, S.F. Matar, M.A. Belkhir and M. Nakhl, Phys. Rev. B {\bf 75}, 064420  (2007). 
\bibitem {13} K. Suzuki, T. Shinohara, F. Wagatsuma, T. Kaneko, H.Yoshida, Y. Obi and S. Tomiyoshi, Jour. Phys. Soc. Jap \textbf{72}, 1175  (2003).
\bibitem {14} H. Shimizu, M. Shirai and N. Zuzuki, J. Phys. Soc. Jap. \textbf{67}, 922 (1998).
\bibitem {15} P.{\nobreakspace}Lukashev and Walter R.{\nobreakspace}L.{\nobreakspace}Lambrecht, 
Phys. Rev. B \textbf{70}, 245205 (2004).
\bibitem {16} S. Matar, L. Fourn\`{e}s, S. Ch\'{e}rubin-Jeannette and G. Demazeau, Eur. J. Solid State Inorg. Chem \textbf{30}, 871 (1993).
\bibitem {17} A. V. dos Santos and C. A. Kuhnen, J. Alloys and Compounds \textbf{321}, 60 (2001).
\bibitem {18} H. P. Meyer and W. Sucksmith, Proc. R. Soc. London, Ser. \textbf{A 207}, 427 (1951).
\bibitem {19} D. Bagayoko, A. Ziegler and J. Callaway, Phys. Rev. B \textbf{27}, 7046 (1983).
\bibitem {20} G. A. Prinz, Phys. Rev. Lett. \textbf{54}, 1051 (1985).
\bibitem {21} C. S. Yoo, H. Cynn, P. Soderlind and V. Iota, Phys. Rev. Lett \textbf{84}, 4132 (2000).
\bibitem {22} V. L. Moruzzi, P. M. Marcus, K. Schwartz and P. Mohn, J. Magn. Magn. Mater \textbf{54-57}, 955 (1986).
\bibitem {23} B. I. Min, T. Oguchi and A. J. Freeman, Phys. Rev. B \textbf{33}, 7852 (1986).
\bibitem {24} J. Donahue, \textit{The structures of the elements} (Wiley, New York, 1976).
\bibitem {25} P. Modak, A. K. Verma, R. S. Rao, B. K. Godwal and R. Jeanloz, Phys. Rev. B \textbf{74}, 01210 (2006).
\bibitem {26} S. Fox and H. J. F. Jansen, Phys. Rev. B \textbf{60}, 4397 (1999). 
\bibitem{27} Z. Ding, P. M. Thibado, C. Awo-Affouda  and V. P. LaBella, J. Vac. Sci. Technol B {\bf 22}, 2068  (2004). 
\bibitem {28} P. Honenberg and W. Kohn, Phys. Rev. 136, B864 (1964); W. Kohn and L.J. Sham, Phys. Rev \textbf{140}, A1133 (1965).
\bibitem {29} J. K\"{u}bler, Phys. Lett. A \textbf{81}, 81 (1981).
\bibitem {30} J. P. Perdew and Y. Wang , Phys. Rev. B \textbf{33}, 8800 (1986).
\bibitem {31} V.\ Eyert, 
{\em The Augmented Spherical Wave Method -- A Comprehensive Treatment}, 
{\em Lecture Notes in Physics} (Springer, Heidelberg, 2007).
\bibitem {32}  J.P. Perdew, S. Burke, and M. Ernzerhof, Phys. Rev. Lett. \textbf{77} 3865 (1996).
\bibitem {33}  R. Hoffmann, Angew. Chem. Int. Ed. Engl. \textbf{26}, 846 (1987). 
\bibitem {34}  R. Dronskowski and P. E. Bl\"{o}chl, J. Phys. Chem. \textbf{97}, 8617 (1993). 
\bibitem{35}  G. Bester and M. F\"{a}hnle, J. Phys.: Cond. Matt. \textbf{13}, 11541 and 11551 (2001).
\bibitem {36}  G. Kresse and J. Hafner, \textit{Phys. Rev. B}, \textbf{47}, 558 (1993);\\
G. Kresse and J. Hafner, \textit{Phys. Rev. B}, \textbf{49}, 14251 (1994);\\
G. Kresse and J. Furthm\"{u}ller, \textit{Comput. Mat. Sci.}, \textbf{6}, 
15 (1996);\\
G. Kresse and J. Furthm\"{u}ller, \textit{Phys. Rev. B}, \textbf{54}, 11169 
(1996).
\bibitem {37}  D. Vanderbilt, \textit{Phys. Rev. B}, \textbf{41}, 7892 (1990).
\bibitem {38}  P.E. Bl\"{o}chl, O. Jepsen, and O. K. Anderson, \textit{Phys. 
Rev. B}, \textbf{49}, 16223 (1994) .
\bibitem {39}  G. Demazeau, A. Wang, S. Matar and  J.D. Cillard. \textit{High Pressure 
Research,} \textbf{12}\textit{,} 343 (1994).
\bibitem {40}  A. D. Becke and K. E. Edgecomb, J. Chem. Phys., \textbf{92}, 5397 (1990).
\bibitem {41} P. Mohn and S.F. Matar, J. Magn. Magn. Mater. \textbf{191}, 234 (1999). 
\bibitem {42} S. F. Matar. J. Alloys and Compounds \textbf{345}, 72 (2002).
\bibitem {43} P. Mohn. {\em Magnetism in the solid State, An Introduction}, (Springer, Berlin, Heidelberg 2003). 
\end{thebibliography}
\end{document}